\documentclass[aps, twocolumn, prx, superscriptaddress, longbibliography]{revtex4-1}
\usepackage{mathptmx} 
\usepackage{amsfonts}
\usepackage{amsmath}
\usepackage{amssymb}
\usepackage{graphicx}
\usepackage{bm}
\usepackage{natbib}
\usepackage{color}
\usepackage[colorlinks=true,linkcolor=blue,anchorcolor=red,citecolor=blue, urlcolor=blue]{hyperref}
\usepackage{comment}

\begin{document}
\title{Gate tuning of fractional quantum Hall states in InAs two-dimensional electron gas}
\author{S.~Komatsu}
\affiliation{Department of Social Design Engineering, National Institute of Technology (KOSEN), Kochi College, 200-1 Otsu Monobe, Nankoku 783-8508 Japan}
\author{H.~Irie}\email{hiroshi.irie.ke@hco.ntt.co.jp}
\affiliation{NTT Basic Research Laboratories, NTT Corporation, 3-1 Morinosato-Wakamiya, Atsugi 243-0198, Japan}
\author{T.~Akiho}
\affiliation{NTT Basic Research Laboratories, NTT Corporation, 3-1 Morinosato-Wakamiya, Atsugi 243-0198, Japan}
\author{T.~Nojima}
\affiliation{Institute for Material Research, Tohoku University, Sendai 980-8577 Japan}
\author{T.~Akazaki}
\affiliation{Department of Social Design Engineering, National Institute of Technology (KOSEN), Kochi College, 200-1 Otsu Monobe, Nankoku 783-8508 Japan}
\author{K.~Muraki}
\affiliation{NTT Basic Research Laboratories, NTT Corporation, 3-1 Morinosato-Wakamiya, Atsugi 243-0198, Japan}
\pacs{PACS number}
\date{\today}

\begin{abstract}
We report the observation of fractional quantum Hall (FQH) effects in a two-dimensional electron gas (2DEG) confined to an InAs/AlGaSb quantum well, using a dual-gated Hall-bar device allowing for the independent control of the vertical electric field and electron density. At a magnetic field of 24 T, we observe FQH states at several filling factors, namely $\nu = 5/3$, $2/3$, and $1/3$, in addition to the $\nu = 4/3$ previously reported for an InAs 2DEG. The $\nu = 4/3$ and $5/3$ states, which are absent at zero back-gate voltage, emerge as the quantum well is made more symmetric by applying a positive back-gate voltage. The dependence of zero-field electron mobility on the quantum-well asymmetry reveals a significant contribution of interface-roughness scattering, with much stronger scattering at the lower InAs/AlGaSb interface. However, the dependence of the visibility of the FQH effects on the quantum-well asymmetry is not entirely consistent with that of mobility, suggesting that a different source of disorder is also at work.
\end{abstract}
\maketitle

\section{Introduction}

The interface between the fractional quantum Hall (FQH) insulator and an \textit{s}-wave superconductor has drawn significant interest as a playground for realizing topological states of matter with non-Abelian quasiparticles \cite{Clarke2013, Clarke2014, Mong2014}. While many theoretical proposals have been made on how to realize such systems, experimental research lags far behind due to the lack of physical systems that fulfill the requirements, namely a clean two-dimensional electron gas (2DEG) hosting FQH states and a transparent 2DEG/superconductor interface allowing for a proximity-induced superconducting phase in the 2DEG. One promising candidate that may fulfill these requirements is graphene; recently, a superconducting proximity effect in FQH edge channels has been demonstrated in graphene edge-contacted with a narrow NbTiN electrode \cite{Gul2021}. Another possible candidate is InAs, which is known to form a transparent contact with superconductors, as demonstrated by the hard superconducting gap in a near-surface InAs layer with an epitaxially grown aluminum film on top \cite{Kjaergaard2016}.
Recent improvements in the quality of InAs/AlGaSb quantum well (QW) structures \cite{Shojaei2016, Tschirky2017, Hatke2017, Thomas2018} have led to the observation of the FQH effect in an InAs 2DEG \cite{Ma2017}.
However, despite the significantly high electron mobility of 180 m$^{2}$/Vs, the only observed FQH state was the one at the Landau-level filling factor $\nu = 4/3$, with no FQH features at other fillings, such as $5/3$, within the same lowest Landau level, where $\nu = nh/eB$ with $n$ the electron density, $h$ Planck's constant, $e$ the elementary charge, and $B$ the magnetic field. It should also be noted that the observation required a high electron density of $7.8 \times 10^{15}$ m$^{-2}$ and thus a high magnetic field of 24 T \cite{Ma2017}.

In this paper, we report the observation of FQH effects at several filling factors---$\nu = 5/3$, $2/3$, and $1/3$ in addition to $4/3$---in an InAs/AlGaSb QW. In particular, we use a dual-gate configuration to tailor the distribution of the electron wave function within the QW and demonstrate that it has a significant impact on the visibility of the FQH effects.
The $\nu = 5/3$, $4/3$ FQH effects, which are absent at zero back-gate voltage ($V_\mathrm{BG}$), emerge as the QW is made more symmetric by applying a positive $V_\mathrm{BG}$ while keeping the electron density the same. We study the correlation between the FQH features and electron mobility as a function of the potential asymmetry. While the results suggest a dominant role of interface-roughness scattering in the mobility variation, it alone cannot account for the behavior of the visibility of the FQH effects, which suggests another source of disorder that affects FQH effects.

\section{Experimental}

\begin{figure}[ptb]
\includegraphics[width = 0.46\textwidth]{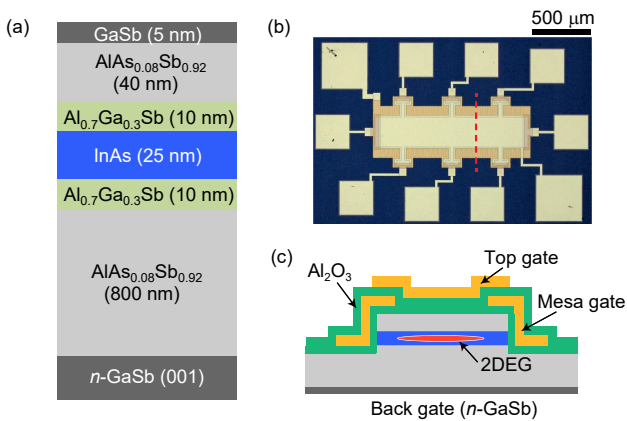}
\caption{\label{Fig1}(color online) (a) Layer structure of the heterostructure. (b) Optical microscope image of the Hall bar. (c) Schematic cross section of the Hall bar along the broken line in (b).}
\end{figure}

We used a 2DEG formed in an InAs/AlGaSb heterostructure grown by molecular beam epitaxy on an \textit{n}-type GaSb (001) substrate. Figure~\ref{Fig1}(a) depicts the layer structure of the wafer. The 2DEG is located in a 25-nm-wide InAs QW sandwiched by 10-nm-thick Al$_{0.7}$Ga$_{0.3}$Sb barriers. The QW structure is flanked on both sides by outer AlAs$_{0.08}$Sb$_{0.92}$ barrier layers and capped with a 5-nm-thick GaSb. The AlAs$_{0.08}$Sb$_{0.92}$ layers were designed to lattice-match the GaSb substrate and thereby suppress dislocation formation.
Despite the absence of intentional doping in the structure, electrons are induced in the QW without gate voltages, mainly due to the Fermi level pinning at the GaSb surface \cite{Nguyen1992}. The electron density and mobility measured in an ungated sample are $4.3 \times 10^{15}$ m$^{-2}$ and 118 m$^{2}$/Vs, respectively. This mobility is comparable to those of recently reported state-of-the-art InAs 2DEGs \cite{Shojaei2016, Tschirky2017, Hatke2017, Thomas2018}, attesting to the high quality of our sample.

The heterostructure was processed into a Hall bar with a gate structure designed specifically for this study as shown in Figs.~\ref{Fig1}(b) and \ref{Fig1}(c). In addition to the global top gate and the back gate (\textit{n}-GaSb substrate), which allow us to independently control the electron density and the electric field across the QW (displacement field), the structure incorporates a third gate (mesa gate), which overlaps only the edge of the mesa and is isolated by atomic-layer-deposited 40-nm-thick Al$_{2}$O$_{3}$ layers [Fig.~\ref{Fig1}(c)].
We electrostatically define the edge of the 2DEG by negatively biasing the mesa gate and depleting the 2DEG underneath \cite{Mittag2018, Mittag2021}. This is because in InAs 2DEGs the Fermi level pinning \cite{Wieder2003} and resultant charge accumulation at sample edges can lead to the formation of counterflowing edge channels and destruction of integer quantum Hall effects \cite{vanWees1995, Akiho2019}.
We note that similar issues are relevant also in graphene \cite{Cui2016, Marguerite2019}, where the use of an edge-free Corbino structure \cite{Zeng2019, Polshyn2018} or an electrostatically defined Hall bar \cite{Ribeiro-Palau2019} has been shown to be effective for the observation of fragile FQH states. To ensure the depletion at the mesa edges, we adjusted the voltage applied to the mesa gate for each $V_\mathrm{BG}$. The distance between adjacent voltage probes and the width of the Hall bar were 400 and 250 $\mu$m, respectively.
Magnetotransport measurements were carried out at magnetic fields of up to 24 T using a cryogen-free superconducting magnet \cite{Awaji2017} in combination with a $^{3}$He refrigerator. The sample was immersed in liquid $^{3}$He with the temperature controlled between 0.48--1.52 K. Some additional diagnostic measurements were performed in a dilution refrigerator (base temperature of 13 mK) equipped with a 12 T superconducting magnet. Longitudinal ($R_{xx}$) and transverse ($R_{xy}$) resistances were measured by standard lock-in techniques with ac excitation current of 100 nA at 23 Hz.

\section{Results and Discussions}

\begin{figure}[ptb]
\includegraphics[width = 0.4\textwidth]{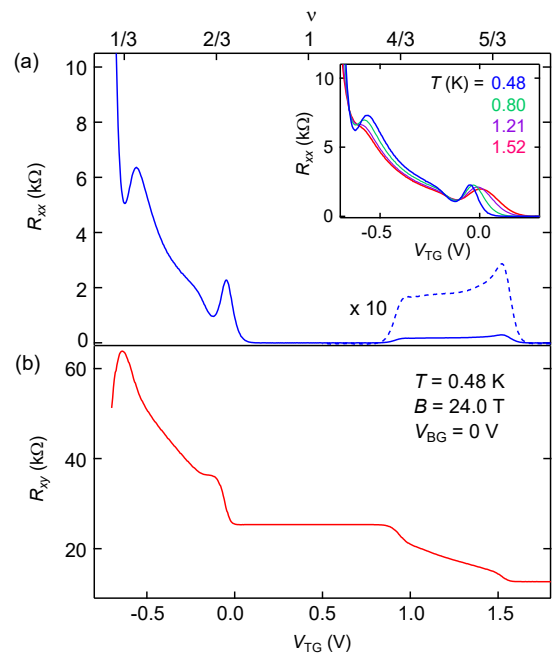}
\caption{\label{Fig2}(color online) $V_\mathrm{TG}$ dependence of $R_{xx}$ (a) and $R_{xy}$ (b) taken at 0.48 K and 24 T with $V_\mathrm{BG} = 0$ V. The broken line in (a) is the tenfold magnification of the data at $\nu > 1$. The inset in (a) shows temperature evolution of $R_{xx}$ traces in the filling factor range including $\nu = 1/3$ and $2/3$.}
\end{figure}

We first show the high-field transport data obtained at $V_\mathrm{BG} = 0$ V. Figures~\ref{Fig2}(a) and \ref{Fig2}(b) depict $R_{xx}$ and $R_{xy}$, respectively, measured at 0.48 K by sweeping the top-gate voltage ($V_\mathrm{TG}$) at $B = 24$ T. We observe deep minima in $R_{xx}$ manifesting the FQH effects, at $V_\mathrm{TG}$'s corresponding to $\nu = 2/3$ and $1/3$ as shown on the top axis.
We also observe a plateau in $R_{xy}$ at $\nu = 2/3$, but not at $\nu = 1/3$, presumably because the 2DEG becomes highly resistive and the Ohmic contacts become poor at low electron density. Interestingly, we see no signatures of FQH states in the region of higher fillings ($1 < \nu < 2$) where the $\nu = 5/3$ and $4/3$ states are expected. This is not trivial, as mobility generally increases with density.

As shown in the inset of Fig.~\ref{Fig2}(a), the $R_{xx}$ minima at $\nu = 2/3$ and $1/3$ are barely temperature dependent for the range studied (0.48--1.52 K). Although not shown, magnetic-field dependent measurements show that the $\nu = 2/3$ FQH effect is visible only at $B > 17$ T, becoming more pronounced with increasing $B$. These results indicate that, at 0.48 K and 24 T, the visibility of the FQH states in our sample is not limited by temperature, but by the disorder present in the sample.

\begin{figure}[ptb]
\includegraphics[width = 0.48\textwidth]{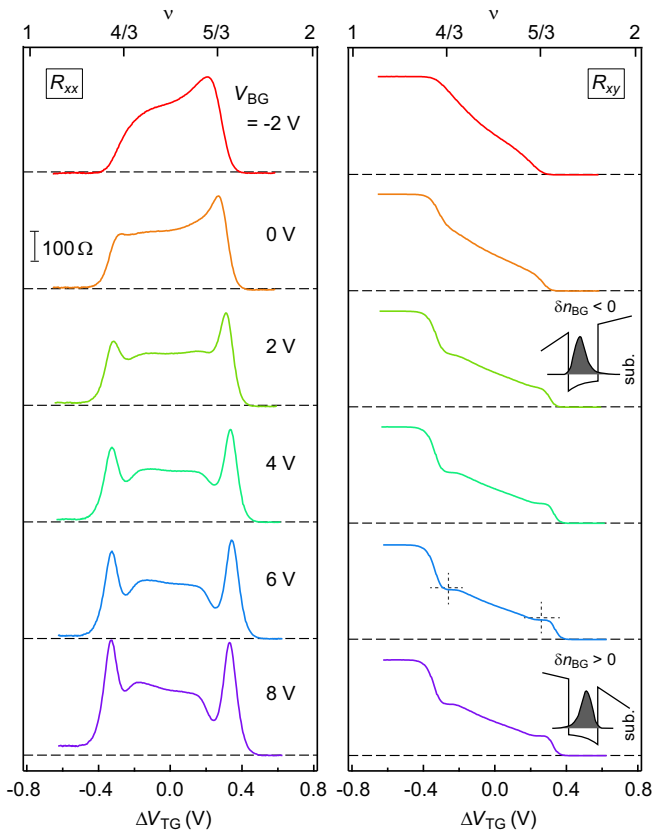}
\caption{\label{Fig3}(color online) $R_{xx}$ and $R_{xy}$ traces for different $V_\mathrm{BG}$ at 0.48 K and 24 T. The horizontal axis, $\Delta V_\mathrm{TG}$, represents the change in $V_\mathrm{TG}$ measured from $\nu = 3/2$ at each $V_\mathrm{BG}$. Each trace is shifted by a constant offset for clarity, with the broken lines indicating the baseline, 0 and $h/2e^{2})$ for $R_{xx}$ and $R_{xy}$, respectively.
The cross marks in the right panel depict the expected positions of the $\nu = 4/3$ and $5/3$ plateaus. The illustrations in the right panel represent the shape of the potential and wave function in the QW for positive and negative $\delta n_\mathrm{BG}$. }
\end{figure}

Now we use the back gate to alter the distribution of the electron wave function within the QW and investigate its influence on the FQH states. For this purpose, below we focus on the filling factor region $1 < \nu < 2$. This is because, at low fillings such as $\nu = 1/3$ and $2/3$, the Ohmic contacts did not work properly when a negative $V_\mathrm{BG}$ was applied.
Figure~\ref{Fig3} shows the $R_{xx}$ and $R_{xy}$ traces which are taken at $B = 24$ T by sweeping $V_\mathrm{TG}$ at different $V_\mathrm{BG}$'s from $- 2$ to 8 V. The data are plotted as a function of $\Delta V_\mathrm{TG}$ (bottom axis) and $\nu$ (top axis), where $\Delta V_\mathrm{TG}$ represents the change in $V_\mathrm{TG}$ measured from $\nu = 3/2$ at each $V_\mathrm{BG}$.
As $V_\mathrm{BG}$ increases, $R_{xx}$ minima emerge at $\nu = 5/3$ and $4/3$, which become more pronounced with increasing $V_\mathrm{BG}$, accompanied by the development of $R_{xy}$ plateaus.

To associate the observed behavior of the FQH effects with the shape of the QW, we need to know at which $V_\mathrm{BG}$ the QW becomes symmetric.
We did this by locating the $V_\mathrm{BG}$ value for which the second subband becomes occupied at the minimum density at a zero magnetic field. Such an analysis reveals that, for $n = 8.7 \times 10^{15}$ m$^{-2}$, which corresponds to $\nu = 3/2$ at 24 T, the QW becomes symmetric at $V_\mathrm{BG} = 4.9$ V.
Thus, we quantify the asymmetry of the QW by $\delta n_\mathrm{BG}$, the electron density added by the back gate with reference to the symmetric point for a given $n$.
By taking into account the electron density supplied from the top gate, we find that the QW becomes symmetric at $V_\mathrm{BG} = 4.2$ and 5.5 V for $\nu = 4/3$ and $5/3$, respectively.
The data in Fig.~\ref{Fig3} suggest that the FQH states appear as the QW is made progressively symmetric.
Interestingly, however, the behavior of the FQH effects is not perfectly symmetric with respect to $\delta n_\mathrm{BG} = 0$.
This can be seen by comparing the data at $V_\mathrm{BG} = 4$ and 8 V for $\nu = 5/3$ and $V_\mathrm{BG} = 2$ and 8 V for $\nu = 4/3$.
In both cases, the $R_{xx}$ minimum is more pronounced for the higher $V_\mathrm{BG}$, i.e., when the wave function is pushed against the lower, rather than the upper, interface despite the greater asymmetry.

\begin{figure}[ptb]
\includegraphics[width = 0.44\textwidth]{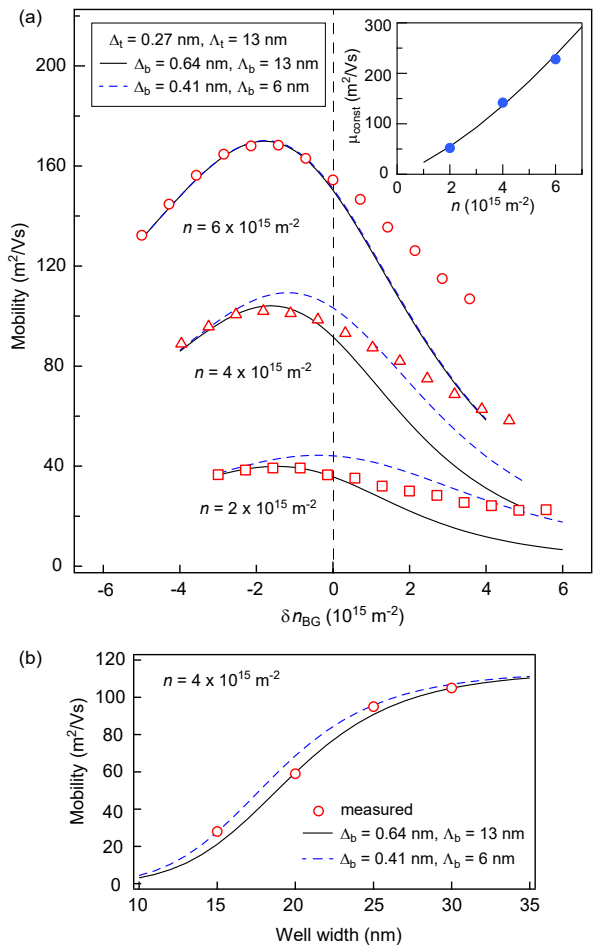}
\caption{\label{Fig4}(color online) (a) Electron mobility as a function of $\delta n_\mathrm{BG}$ for constant electron densities of $2 \times 10^{15}$ (square), $4 \times 10^{15}$ (triangle), and $6 \times 10^{15}$ m$^{-2}$(circle). The solid and broken lines represent calculated total mobility taking into account interface-roughness scattering (for two sets of parameters shown in the figure) and $\delta n_\mathrm{BG}$-independent constant term, the latter being a fit parameter.
The closed circles in the inset indicate this $\delta n_\mathrm{BG}$-independent constant term used for each $n$. The solid line represents calculated mobility limited by background impurities with the concentration of $n_\mathrm{BI} = 5.2 \times 10^{15}$ cm$^{-3}$. (b) Electron mobility at $n = 4 \times 10^{15}$ m$^{-2}$ for samples with different QW thicknesses. The solid and broken lines are mobility calculated using the same parameters as in (a) for the interface roughness and background impurity concentration.}
\end{figure}

To identify the source of disorder that limits the visibility of the FQH states, we examine the behavior of electron mobility, the most common measure of sample quality.
In Fig.~\ref{Fig4}(a), electron mobility for the constant densities of 2, 4, and $6 \times 10^{15}$ m$^{-2}$ is plotted as a function of $\delta n_\mathrm{BG}$.
The data show that mobility drops significantly when a large $|\delta n_\mathrm{BG}|$ is imposed and the wave function is pushed against the upper or lower InAs/AlGaSb interface.
Such a rapid mobility drop is not expected for long-range Coulomb scattering by localized charges. It is also noteworthy that the mobility variation is not symmetric with respect to $\delta n_\mathrm{BG} = 0$.
The strong $\delta n_\mathrm{BG}$ dependence suggests that short-range scattering by interface roughness is the dominant contributor.
The asymmetry with respect to $\delta n_\mathrm{BG} = 0$ can then be accounted for by the difference between the upper and lower interfaces.

We have calculated mobility limited by interface-roughness scattering following the method in Ref. \cite{Ando1982}. For both interfaces, we started with the parameters reported for InAs/AlGaSb QWs \cite{Shojaei2016}, roughness height of $\Delta = 0.27$ nm and correlation length of $\Lambda = 13$ nm.
We then find that, by increasing the roughness height of the lower interface to $\Delta_\mathrm{b} = 0.64$ nm while keeping other parameters the same, the observed asymmetry can be reproduced reasonably well [Fig.~\ref{Fig4}(a)]. To fit the $\delta n_\mathrm{BG}$ dependence, one needs to assume an additional scattering mechanism that does not depend on $\delta n_\mathrm{BG}$.
This $\delta n_\mathrm{BG}$-independent contribution used for the fit is plotted in the inset as a function of density. These values agree well with the calculation of mobility limited by background impurities assuming an impurity concentration of $n_\mathrm{BI} = 5.2 \times 10^{15}$ cm$^{-3}$, corroborating that the $\delta n_\mathrm{BG}$-independent contribution comes from background-impurity scattering.
Calculation using the same interface parameters and $n_\mathrm{BI}$ reproduces the dependence of mobility on the QW thickness measured for different samples [Fig.~\ref{Fig4}(b)].

For the fit shown in Fig.~\ref{Fig4}(a), we chose the parameters to reproduce the behavior near the mobility peak, which unavoidably resulted in the underestimation of mobility at large $\delta n_\mathrm{BG}$. This remained true when the correlation length for the lower interface was adjusted as an additional parameter [dashed lines in Fig.~\ref{Fig4}(a)].
It is known that Sb segregation during growth results in the incorporation of Sb atoms into the InAs layer and causes alloy scattering. Although this could be another source of the asymmetry between positive and negative $\delta n_\mathrm{BG}$, we have confirmed that the influence of alloy scattering, calculated for the composition profile reported in Ref. \cite{Steinshnider2000}, is negligibly small as compared to interface-roughness scattering.
We add that the quantum lifetime evaluated from the low-field Shubnikov-de Haas oscillations was around 0.5 ps, with no significant $V_\mathrm{BG}$ dependence.

We note that the behavior of the FQH effects is not entirely consistent with the $\delta n_\mathrm{BG}$ dependence of mobility. For example, upon increasing $\delta n_\mathrm{BG}$ from $-2.1 \times 10^{15}$ to 0 m$^{-2}$, mobility decreases, whereas the FQH states become better developed. [For $\nu = 4/3$ ($5/3$), $\delta n_\mathrm{BG} = -2.1 \times 10^{15}$ and 0 m$^{-2}$ correspond to $V_\mathrm{BG} = 1.2$ (2.6) and 4.2 (5.5) V, respectively.]
It is known that short-range disorder such as alloy disorder is less detrimental than long-range disorder such as Coulomb scattering on FQH effects \cite{Deng2014,Pan2011,Betthausen2014,Kleinbaum2020}. Since interface-roughness scattering is short-ranged in nature, it is not surprising that the stronger roughness scattering on the lower interface does not significantly deteriorate the visibility of the FQH effects.
However, it leaves the question of why the FQH effects disappear when the wave function is pushed strongly against the upper interface. One possibility is the potential fluctuation arising from the localized charges trapped at the GaSb/Al$_{2}$O$_{3}$ interface.
It is known that 2DEG is supplied from the surface states at the semiconductor surface. Our calculation shows that the center of the electron wave function shifts by 6 nm over the range of $V_\mathrm{BG}$ in Fig.~\ref{Fig3}. This shift is not negligible as compared to the distance from the upper InAs/AlGaSb interface to the GaSb/Al$_{2}$O$_{3}$ interface (55 nm).
This implies that the fluctuation in the long-range Coulomb potential that the 2DEG experiences becomes weaker when the wave function shifts toward the lower interface. Although the influence of surface charges on mobility is minor, their impact on FQH effects may not be negligible.
This is similar to the case of remote ionized impurities in modulation-doped GaAs/AlGaAs heterostructures or QWs, where their impact on FQH effects is significant even when their influence on mobility is minor \cite{Gamez2013}.

It is worth mentioning that the InAs 2DEG samples used in most studies, including ours, are undoped, where the electrons are induced by the Fermi-level pinning at the semiconductor surface \cite{Nguyen1992}. Our study suggests that the surface charge may affect the visibility of the FQH effects, which could be a reason why FQH effects are not readily observable in InAs 2DEGs in spite of the high electron mobility.
One way to mitigate this might be to increase the distance from the surface to the 2DEG. This would inevitably reduce the electron density, which could neverthless be compensated by using a front or back gate as we partially demonstrated here. It should be also mentioned that in this study we used a specially designed Hall bar, in which the edges of the 2DEG were electrostatically defined by gates.
If the physical edges of the sample are to be used, one would need to take care to avoid the influence from the possible charge accumulation at the edges \cite{vanWees1995,Akiho2019}, which may affect FQH effects. Lastly, we must note that the high mobility in InAs/AlGaSb QWs is partly due to the small effective mass in InAs; even in high-quality samples, the background impurity concentration (a few 10$^{15}$ cm$^{-3}$) is still much higher than in GaAs/AlGaAs samples of moderate quality (a few 10$^{14}$ cm$^{-3}$). It is therefore essential to further improve the crystal quality to realize FQH effects at lower magnetic fields compatible with superconductivity.

\section{Summary}

We have observed FQH effects at $\nu = 5/3$, $4/3$, $2/3$, and $1/3$ in an InAs 2DEG using a dual-gated Hall bar with gate-defined edges. We demonstrated that the shape of the QW confinement potential has a significant impact on the visibility of the FQH states. Comparison with the variation of zero-field electron mobility suggests an additional source of disorder, other than interface roughness and background impurities, that affects the FQH effects. Our results hint at the need to further optimize the heterostructure design, in addition to improving crystal quality, for the realization of robust FQH states in InAs.

\section*{Acknowledgments}

The authors thank Hiroaki Murofushi for device fabrication, Miki Imai for constructing measurement setups, and Norio Kumada for discussions and useful comments on the manuscript. High-field measurements were performed at the High Field Laboratory for Superconducting Materials, Institute for Materials Research, Tohoku University (Project No. 20H0066).

S.K. and H.I. contributed equally to this work.


%

\end{document}